\documentclass[amsmath,aps,showpacs,preprint]{revtex4}
\begin{document}

\title{Mean Free Path in Disordered Multichannel Tight-binding Wires }
\author{Jean Heinrichs}
\email{J.Heinrichs@ulg.ac.be} \affiliation{D\'{e}partement de
Physique, B{5a}, Universit\'{e} de Li\`{e}ge, Sart Tilman,
B-{4000} Li\`{e}ge, Belgium}
\date{\today}

\begin{abstract}
Transport in a disordered tight-binding wire involves a collection
of different mean free paths resulting from the distinct fermi
points, which correspond to the various scattering channels of the
wire. The generalization of Thouless' relation between the mean
free path and the localization length $\xi$ permits to define an
average channel mean free path,$\bar\ell$, such that $\xi\sim
N\bar\ell$ in an $N$-channel system.  The averaged mean free path
$\bar\ell$ is expressed exactly in terms of the total reflection
coefficient of the wire and compared with the mean free path
defined in the maximum entropy approach.
\end{abstract}


\maketitle

\section{INTRODUCTION}

In a series of papers\cite{1,2,3} (hereafter referred to as I, II,
III) we studied localization for weak disorder in few channel
($N$) wires, using a scattering matrix approach.  The disordered
wires in I and II were described by the Anderson tight-binding
model and two- and three chain systems were considered where, for
$N=3$, we further distinguished between chains arranged on a
"tube" (periodic lateral boundary conditions) or on a "strip"
(free lateral boundary conditions).  The multichannel systems of
length $N_L$ (in units of the lattice parameter $a$) were
connected as usual to non-disordered leads.

In I we restricted to the familiar case where all channels are
conducting at the fermi level.  In II we extended our treatment to
the case where one or several states at the fermi level are
evanescent.  In III we applied our treatment in the case of a
model involving equivalent tight-binding chains with both random
site energies and random interchain hoppings\cite{4}.

In addition to the localization length, $\xi$,we also studied
elastic mean free paths, $\ell$,  in II and III, using our
scattering matrix results for averaged channel reflection
coefficients.  We used an expression for the mean free path in
terms of the total reflection coefficient defined in the framework
of a maximum entropy approach to multichannel conductors\cite{5}.
The maximum entropy approach has been further discussed in refs.
6,7 and has received considerable development as shown in
extensive reviews\cite{8,9,10,11}.  Unfortunately the mean free
paths calculated from the expression proposed in
\cite{5,6,7,8,9,10,11}, using our reflection
coefficients\cite{1,2,3}, led to a very unsatisfactory general
result\cite{2,3}, namely

\begin{equation}\label{eq1}
\ell=\frac{\xi}{2}\quad ,
\end{equation}
for all several-channel systems (and  for various positions of the
fermi level in each case) studied in I-III.  Here $\xi$ refers to
the exact weak disorder localization lengths obtained from the
conductance in various cases\cite{1,2,3}.  The main problem of
(\ref{eq1}) is of course that its extrapo\-lation to the many
channel case would imply the non-existence of a mesoscopic
metallic regime in a thick wire, in contrast to well-known
results\cite{12,13}. This reveals an important insufficiency of
the expression for the mean free path used as a constraint in the
maximum entropy approach.

The purpose of this paper is to derive a new general expression of
the mean free path which is rooted in the microscopic aspects of
tight-binding wires such as those studied in I-III.  The
microscopic description leads us to define different mean free
paths for the various conducting channels in a wire.  From the
individual channel mean free paths one may define an average
channel mean free path which, in turn may be expressed in terms of
the total reflection coefficient of the disordered wire.  The use
of this average mean free path expression palliates to the
unsatisfactory features encountered with the mean free path
expression of the maximum entropy approach.  This is discussed in
Sect. II below.

\section{MEAN FREE PATH FORMULA}

We first recall some relevant aspects of the study of the coupled
tight-binding chain models of wires in I. In a first step of the
analysis\cite{1} the interchain hopping has been diagonalized in
order to define new independent chains for the non-disordered
leads.  The Bloch wave solutions for these individual chains then
define a set of independent channels for wave propagation and
scattering by the disorder.The Bloch energy band associated with a
channel $i$ in any of the few-channel systems studied in I is of
the form

\begin{equation}\label{eq2}
E_i=\alpha_i-2\beta\cos k_i\quad,
\end{equation}
where $-\beta$ is a constant hopping parameter between
nearest-neighbour sites on the chain and $\alpha_i$ is related to
a constant nearest-neighbour interchain hopping rate $h$.
Specifically we have

\begin{equation}\label{eq3}
\alpha_1=0
\end{equation}
for a single-chain (one-dimensional) system,

\begin{equation}\label{eq4}
\alpha_1=h\quad,\quad\alpha_2=-h\quad,
\end{equation}
for a two-chain system ($i=1,2$),

\begin{align}\label{eq567}
    \alpha_1&=\sqrt{2}h\\
    \alpha_2&=0\\
      \alpha_3&=-\sqrt{2}h\quad ,
\end{align}
for a three-chain \textbf{strip} ($i=1,2,3$) and, finally,

\begin{align}\label{eq89}
\alpha_1&=2h\\
 \alpha_2&=\alpha_3=-h\quad ,
\end{align}
for a three-chain \textbf{tube} ($i=1,2,3$).

Clearly for a wire composed of an arbitrary number $N$ of
tight-binding chains the channel-basis band structure will include
$N$ one-dimensional energy bands of the form(\ref{eq2}).  The
allowed Bloch wavenumber values obtained by applying Born-von
Karman boundary conditions are

\begin{equation}\label{eq10}
k_i=\frac{2\pi m}{L}, m=0,1,2,\ldots\quad ,
\end{equation}
where $L$ is the length of the wire.

In studying electrical transport in an $N$-channel wire with the
band structure above we focus on typical situations where bands of
the form(\ref{eq2}) are mutually overlapping so that e.g. a number
$N_c\leq N$ of the bands are conducting at the fermi level of
energy $E_F$.  This implies that the conduction electrons at the
fermi level are distributed over $N_c$ distinct energy bands
(\ref{eq2})at corresponding fermi points.

In this case the Drude conductivity $\sigma$ of the wire is given
by\cite{14}

\begin{equation}\label{eq11}
\sigma= e^2 \sum^{N_c}_{i=1}\frac{n_i\tau_i}{m_i}\quad ,
\end{equation}
where $n_i$ is the electron density in the $i$th channel, $\tau_i$
is the relaxation time for electrons in channel $i$, and $m_i$ is
the effective mass evaluated at the fermi level ,

\begin{equation}\label{eq12}
m_i=\frac{1}{\hbar^2}\frac{\partial^2 E_i}{\partial
k^2_i}\biggl\lvert_{E_i=E_F}\quad .
\end{equation}
We further define the electron velocity in the $i$th branch, at
the fermi level

\begin{equation}\label{eq13}
v_{F_i}=\frac{1}{\hbar}\frac{\partial E_i}{\partial
k_i}\biggl\lvert_{E_i=E_F}\quad .
\end{equation}
Now $n_i$ is given by $n_i=\frac{N_i}{AL}$ where $N_i$ is the
total number of electrons in the band $E_i$ in the ground state
and $A$ denotes the cross section area of the wire: $N_i$ is
obtained by filling the $k$-levels with two electrons each up to
the fermi level at $k_{F_i}$ given by

\begin{equation}\label{eq14}
\cos k_{F_i}=\frac{E_F-\alpha_i}{2\beta}\quad .
\end{equation}
From (\ref{eq10} we obtain $N_i=k_{F_i}L/\pi$.  By inserting these
results in (\ref{eq11}) and defining the mean free path  in the
$i$th channel, $\ell_i=v_{F_i}\tau_i$ we finally get

\begin{equation}\label{eq15}
\sigma=\frac{2 e^2}{h A}\sum^{N_c}_{i=1}\ell_i
\end{equation}

In mesoscopic systems one is primarily interested in the
sample-specific conductance which, in the metallic regime to be
identified, is given by Ohm's law, $g=\sigma A/L$.  Using
(\ref{eq15}) we thus obtain

\begin{equation}\label{eq16}
gL=\frac{2e^2}{h}\sum_i \ell_i\quad .
\end{equation}
Now, as is well-known\cite{15}, the order of magnitude of the
localization length $\xi$ is given by the range of lengths where
the reduced or dimensionless conductance $g/(2e^2/h)$ is of order
one, since beyond that range the conductance decreases drastically
(insulating regime).  Thus from (\ref{eq16}) we obtain the
important result

\begin{equation}\label{eq17}
\xi\sim\sum^{N_c}_{i=1}\ell_i\quad ,
\end{equation}
which proves the additivity of the localization length with
respect to mean free paths in the individual channels.  This shows
in addition that the mesoscopic metallic domain extending over
length scales from a typical mean free path up to the localization
length will be the wider the larger the number of channels i.e.
the thicker the wire.

It may be useful in practice to define an average mean free path
over the fermi surface,

\begin{equation}\label{eq18}
\bar\ell=\frac{1}{N_c}\sum_i\ell_i\quad ,
\end{equation}
of magnitude

\begin{equation}\label{eq19}
\bar\ell=\frac{\xi}{N_c}\quad ,
\end{equation}
where we recall that $N_c\leq N$ is the number of conducting
channels at the fermi level.  In this case the range for the
metallic domain reads

\begin{equation}\label{eq20}
\bar\ell\ll l\ll N_c\bar\ell =\xi
\end{equation}
The use of a single elastic mean free path has been a common
practice in the study of multichannel wires, starting with the
influential paper of Thouless\cite{12} on the proof of
localization in such wires.  In fact our analysis generalizes the
discussion of Thouless in the case of a multichain tight-binding
wire where the mean free paths in the conducting channels are
generally different.

Finally, besides the definition (\ref{eq19}) of an average mean
free path, it is useful to obtain an exact expression of it in
terms of scattering parameters, in analogy with a definition used
in the maximum entropy approach\cite{5,6,7,8,9,10,11}.  We now
derive such an expression for weak disorder.  From (\ref{eq19})
and the definition of the localization length from conductance
(\ref{eq1}) we have

\begin{equation}\label{eq21}
\frac{1}{N_c\bar\ell}=-\lim_{L\rightarrow\infty}\frac{\langle\ln
g\rangle}{2L}\quad ,
\end{equation}
where $g$ is the Landauer two-probe conductance

\begin{equation}\label{eq22}
g=\frac{2e^2}{h}Tr(\hat t\hat t^+)\quad ,
\end{equation}
$\hat t$ is the transmission amplitude matrix of the wire and
$\langle\ldots\rangle$ means averaging over the disorder.  From
current conservation in an $N$-channel wire\cite{2} it follow that

\begin{equation}\label{eq23}
Tr\hat t\hat t^+=N-Tr\hat r\hat r^+\quad ,
\end{equation}
where $\hat r$ is the reflection amplitudes matrix.  For weak
disorder transmission coefficients are close to 1 and reflection
coefficients are close to zero.  Thus from (\ref{eq21}-\ref{eq23})
we obtain the exact expression

\begin{equation}\label{eq24}
\frac{1}{\bar\ell}=\frac{\langle Tr(\hat r\hat
r^+)\rangle}{2L}\quad ,
\end{equation}
for weak disorder and sufficiently large $L$\cite{16}.

The essential difference between (\ref{eq24}) and the expression
for $1/\ell$ used in the maximum entropy
approach\cite{5,6,7,8,9,10,11} lies in the presence in the latter
of an additional factor of $1/N$ (see e.g. (3.10) in Ref. 5).
Returning to the detailed calculation of localization lengths and
mean free paths in few channel systems in I-III we note that by
replacing the mean free paths $\ell$ based on the maximum entropy
expression (Eq. (57) in II) by the present correct result,
$\bar\ell=\ell/N_c$, we obtain from (\ref{eq1}),
$\xi=2N_c\bar\ell$ for all cases analyzed in I-III. This finally
confirms that the coupled disordered tight-binding chain model
provides a sound microscopic basis for describing multichannel
wires and the corresponding mesoscopic metallic domain of these
systems.

\end{document}